\documentclass[a4paper,14pt]{extreport}
\usepackage{cmap} 
\usepackage[cp1251]{inputenc}
\usepackage[english,russian]{babel}

\usepackage{braket,amssymb,amsmath,array,commath,mathtext,wrapfig,tikz}
\usepackage[arrows=pgf-filled,version=3]{mhchem}
\usepackage{bm}
\usepackage{amsthm,amsfonts,amscd}
\usepackage{graphicx}
\usepackage{cite}
\usepackage{lscape}
\usepackage{geometry}
\usepackage{indentfirst}
\usepackage{afterpage,soul,ulem,dcolumn,changebar,longtable,hhline,multirow,makecell}
\usepackage[small,bf]{caption}
\addto\captionsrussian{%
}

\makeatletter
\renewcommand{\@biblabel}[1]{#1.\hfil} 
\makeatother

\begin{document}
\textit{Non-linearity of Vaidya spacetime and forces in the central naked singularity} \\
\textbf{Vitalii Vertogradov} \\

\textbf{Affiliation:} Herzen State Pedagogical University of Russia, Moika 48, St. Petersburg 191186, Russia \\ The SAO RAS, Pulkovskoe shosse 65, 196140, Saint-Petersburg, Russia \\

\textbf{e-mail} vdvertogradov@gmail.com \\

\textbf{Summary.} In this paper we consider non-linear Vaidya spacetime i.e. the case when the mass function has the non-linear form $M(v) \equiv \lambda v^n \,, \lambda >0 \,, n>1 $. We prove that the central naked singularity might form for values $n>1$ but they are gravitationally weak. Also we investigate the question about  forces in the naked singularity and prove that they might be finite only in the case of the gravitationally weak naked singularity.

\textbf{Key words.} Vaidya spacetime, naked singularity, strength of singularity.

UDC                   524.8

\setcounter{equation}{0}
\section{1. Introduction}

The Vaidya spacetime is one of the earliest counterexamples of cosmic censorship conjecture [CCC] violation ~\cite{bib:pap}. Vaidya spacetime is so-called radiating Schwarzschild metric with non-zero right hand-side of the Einstein equation. The energy-momentum tensor in Vaidya spacetime represents the null dust and has the following form:

\begin{equation}
T_{ik}=\rho \delta^0_i\delta^0_k \,,
\end{equation}
where $\rho$ is the energy density of this null dust.

If we consider the gravitational collapse of usual Vaidya spacetime~\cite{bib:jv} then the result of such a process might be the naked singularity formation. It means that there exist a family of non-spacelike future-directed geodesics which terminate in the central singularity in the past. Moreover, the time of the singularity formation must be less than the time of the apparent horizon formation.  The geodesic motion represents the free  movement. It means there is no any non-gravitational force acting on the particle. However, there are so-called inertial forces and the particle moves in the medium because this spacetime is filled with the null dust.  We can calculate these forces and, as we see below, in the case of linear Vaidya spacetime  (The mass function is the linear one) these forces are infinite. We will show that in the case of non-linear Vaidya spacetime these forces might be finite. 

We will show that the naked singularity which is formed in the case of non-linear Vaidya spacetime is the gravitationally weak. If we follow Tipler definition  which was given in the paper~\cite{bib:ir}: a singularity is termed  to be gravitationally strong or simply strong if it destroys by stretching  or  crushing any object which falls into it. If it does not destroy any object this way then the singularity is termed to be gravitationally weak. 

The purpose of this paper is to investigate the inertial forces in the case of the naked singularity formation and to find out when these forces might be finite. Also the question about the strength of the singularity is considered.

This paper is organized as follows: in sec. II we consider the possibility of the naked singularity formation in the case $M(v)=\lambda v^n$, in sec. III we consider the inertial forces in Vaidya spacetime. In sec. IV we investigate the question about the strength of the naked singularity in the case of non-linear Vaidya metric. Sec. V is the conclusion.

Throughout the paper the system of units $c=G=1$ will be used.

\setcounter{equation}{0}
\section{2. Naked singularity formation}

The usuall Vaidya spacetime has the following form:

\begin{equation}
ds^2=-\left ( 1-\frac{2M(v)}{r} \right ) dv^2+2\varepsilon dv dr +r^2d\Omega^2 \,,
\end{equation}

where $\varepsilon =\pm 1$ - ingoing (outgoing) radiation, $M(v)$ - the mass function, $v_{\pm 1}$ - advanced (retarded) Edington's time.  $d\Omega$ is the metric on the unit two sphere:
\begin{equation}
d\Omega^2\equiv d\theta^2+\sin^2\theta d\varphi^2 \,.
\end{equation}

We are interested in the gravitational collapse case and because of this we will put $\varepsilon =+1$ throughout  the paper.

\subsection*{The case $M(v)=\lambda v^n$}

Let's consider the case when the mass function has the form:

\begin{equation}
M(v)=\lambda v^n \,,
\end{equation}

where $\lambda$ - a positive real constant.

The equation of the apparent horizon is given by~\cite{bib:pois}:

\begin{equation}
g_{00}=\frac{2 \lambda v^n}{r}-1 =0 \,.
\end{equation}

We can see that if $v=0$ then we can't satisfy the apparent horizon equation (we are interested in positive values of $r$). 

At the time $v=0$ and at the point $r=0$ we have the singularity formation but to be the naked one there must be non-spacelike future-directed geodesics which terminate at the central singularity in the past. Let's consider the existence of radial null geodesics.  The geodesic equation in this case is given by:

\begin{equation}
\frac{dv}{dr}=\frac{2r}{r-2\lambda v^n} \,.
\end{equation}

the geodesic can originate from the central singularity if $\lim\limits_{v\rightarrow 0, r\rightarrow0}\frac{dv}{dr}=X_0$ where $X_0$ - a finite positive number.

We should consider 3 cases
\begin{itemize}
\item $0<N<1$ ,
\item $n=1$,
\item $n>1$.
\end{itemize}

The case of linear mass function $n=1$ has been considered in~\cite{bib:jv}. In this case we have the naked singularity formation if $\lambda < \frac{1}{8}$. Also this singularity is gravitationally strong~\cite{bib:ir, bib:tip}.

Let's consider the case $n<1$. Let's denote $\lim\limits_{v\rightarrow 0, r\rightarrow0} \frac{dv}{dr}=X_0$ then

\begin{equation}
\begin{split}
X_0=\frac{2}{1-2\lambda n X_0v^{n-1}} \,, \\
2\lambda n v^{n-1}X_0^2-X_0+2=0 \,.
\end{split}
\end{equation}

$v\rightarrow 0$ and $n<1$ hence   from the last equation we can see $X_0$ can't be positive real constant and $X_0 \rightarrow\infty $. And we can conclude that there are  no radial null geodesics which terminate at the central singularity in the past in this case.

Now let's consider the case when $n>1$. In this case we have:

\begin{equation} \label{eq:geo2}
\begin{split}
X_0=\frac{2}{1-2\lambda n X_0v^{n-1}} \,, \\
2\lambda n v^{n-1}X_0^2-X_0+2=0 \,.
\end{split}
\end{equation}

Here we have three possible cases:
\begin{itemize}
\item $X_0=0$. In this case we don't have any uncertainty and the geodesic equation \eqref{eq:geo2} gives impossible equality. So this case is impossible.
\item $X_0 \rightarrow\infty$. In this case the condition $\lim\limits_{r\rightarrow0\, v\rightarrow0}v^{n-1}X_0=\frac{1}{2\lambda n}$ must be held. But again the last equation \eqref{eq:geo2} gives impossible equality and we can conclude that this case don't suit us.
\item $X_0=\mu$, where $\mu$ is real positive constant. In this case we have an equality $X_0=2$ and only this case suits us.
\end{itemize}

Note that in this case we have a radial null geodesic which terminates at the central singularity in the past regardless of value $\lambda$. Further we will show that in the case $n>1$ this singularity is gravitationally weak.

\setcounter{equation}{0}
\section{3. The inertial forces}

To investigate the question about the inertial forces one should consider the second order geodesic equation. In Vaidya spacetime they are given by:

\begin{equation}
\begin{split}
\frac{d^2 t}{d\tau^2}=-\frac{M(v)}{r^2}\left ( \frac{dt}{d\tau} \right ) ^2 +r\left ( \frac{d\theta}{d\tau} \right )^2+r\sin^2\theta \left (\frac{d\varphi}{d\tau}\right )^2 \,.
\end{split}
\end{equation}

\begin{equation}
\begin{split}
\frac{d^2 r}{d\tau}=-\frac{\left (1-\frac{2M(v)}{r}\right ) M(v)+\dot{M}(v)r}{r^2}\left (\frac{dt}{d\tau} \right )^2 +2\frac{M(v)}{r^2}\frac{dt}{d\tau}\frac{dr}{d\tau} \\
+\left (r-2M(v)\right )\left (\frac{d\theta}{d\tau} \right )^2+\left (r-2M(v)\right ) \sin^2\theta \left (\frac{d\varphi}{d\tau} \right )^2 \,.
\end{split}
\end{equation}

\begin{equation}
\begin{split}
\frac{d^2\theta}{d\tau^2}=-\frac{2}{r}\frac{dr}{d\tau}\frac{d\theta}{d\tau}+\sin\theta\cos\theta\left(\frac{d\varphi}{d\tau} \right )^2 \,.
\end{split}
\end{equation}

\begin{equation}
\begin{split}
\frac{d^2\varphi}{d\tau^2}=-\frac{2}{r}\frac{dr}{d\tau}\frac{d\varphi}{d\tau}-2\cot \theta\frac{d\theta}{d\tau}\frac{d\varphi}{d\tau} \,.
\end{split}
\end{equation}

One should note that only the following combination of four-velocity $u^0u^0$ and $u^0u^\alpha \,, \alpha =1\,, 2\,, 3$ give us the inertial forces. The combination $u^\alpha u^\beta$ is the part of three covariant derivative and they are not forces at all~\cite{bib:landau}.

Here we have 2 different forces $\Gamma^0_{00}=-\Gamma^1_{01}=\frac{M(v)}{r^2} $ and $\Gamma^1_{00}=\frac{\left ( 1 -\frac{2M(v)}{r} \right ) M(v)+\dot{M}(v)r}{r^2}$. We are interested in the case of the naked singularity formation. As we found out above in the case of the mass function $M(v)=\lambda v^n$ the result of the gravitational collapse is the naked singularity when $n\geq 1$. Using the well-known formula~\cite{bib:landau}:

\begin{equation} \label{eq:force}
\begin{split}
u^\alpha=\frac{1}{\sqrt{1-\beta^2}}\frac{dx^\alpha}{dt} \,, \\
u^0=\frac{1}{\sqrt{1-\frac{2\lambda v^n}{r}}\left (1-\beta^2\right ) }-\frac{r}{(r-2\lambda v^n)(1-\beta^2)}\frac{dx^\alpha}{dt} \,,
\end{split}
\end{equation}

we can easily obtain the radial force  expression:

\begin{equation}
\begin{split}
F^r_{cent}=-\frac{\left (1-\frac{2\lambda v^n}{r} \right) \lambda v^n +\lambda nv^{n-1}r}{r^2\left (1-\frac{2\lambda v^n}{r} \right ) \left (1-\beta^2 \right )} \,, \\
F^r_{nc}=\frac{2}{\sqrt{1-\frac{2\lambda v^n}{r}}\left ( 1-\beta^2 \right)} \left [\frac{\left (1-\frac{2\lambda v^n}{r}\right) \lambda v^n+\lambda n r v^{n-1}}{r^2}-\frac{\lambda v^n}{r^2} \right ]\frac{dr}{dt} \,.
\end{split}
\end{equation}

Where $F_{cent}$ is the centrifugal force and $F_{nc}$ is the force which depends upon a velocity $\frac{dr}{dt}$ linearly. We know that when $n\geq 1$ then $\lim\limits_{r\rightarrow0\,, v\rightarrow0}\frac{dv}{dr}=X_0$ where $X_0$ is a positive real constant. Thus to find the forces in the central naked singularity one should consider three cases:

\begin{enumerate}
\item $n \in [1\,, 2)$. In this case it is easy to see that  we have infinity inertial forces:

\begin{equation}
\begin{split}
\lim\limits_{r\rightarrow0\,, v\rightarrow0} F^r_{cent}=\infty \,, \\
\lim\limits_{r\to 0\,, v\to 0}F^r_{nc}=\infty \,.
\end{split}
\end{equation}

\item $n=2$. In this case we have:
\begin{equation}
\begin{split}
\lim\limits_{r\rightarrow0\,, v\rightarrow0}F_{cent}^r=-\frac{1}{1-\beta^2} \left ( (\lambda X_0^2+2\lambda X_0 \right) \,, \\
\lim\limits_{v\rightarrow0\,, r\rightarrow0}F^r_{ns}=\frac{2}{1-\beta^2} 2\lambda X_0\frac{dr}{dt} \,. 
\end{split}
\end{equation}

We can see that in this case the radial inertial forces have a real positive finite value.
\item $N>2$. In this case it is easy to see that all inertial forces are equall to zero.
\end{enumerate}

\setcounter{equation}{0}
\section{4. The strength of the central singularity}

According  to the Tipler's definition the singularity is strong if the following condition is held~\cite{bib:tip, bib:ir}:

\begin{equation} \label{eq:ric}
\lim\limits_{\lambda \rightarrow 0} \lambda^2R_{ik}K^iK^k=\xi> 0 \,,
\end{equation}

where $\lambda$ is affine parameter, $K^i$ is the tangent vector to the singularity, $R_{ik}$ is the Ricci tensor. Also $\xi$ must be finite. In the case of usual Vaidya spacetime the condition \ref{eq:ric} gives for the expression $\xi$~{bib:mah}:

\begin{equation}
\xi=\frac{2\dot M(v)}{r^2}v^2 \,.
\end{equation}

We are interested in the case $M(v)=\lambda v^n$. One can obtain:

\begin{equation}
\xi=\frac{2nv^{n+1}}{r^2} \,.
\end{equation}

Now we should consider the following  limit (As in the previous cases we denote $\lim\limits_{v\rightarrow0, r\rightarrow0} \frac{dv}{dr}=X_0$:

\begin{equation}
\lim\limits_{v\rightarrow0} \xi= 2nv^{n-1}x_0^2 \,,
\end{equation}

From this we can conclude that if $n=1$ then we have gravitationally strong naked singularity. If $n>1$ then $\xi=0$ and we have only gravitationally weak naked singularity.

From this result we can conclude that in Vaidya spacetime in the case of the naked singularity formation the inertial forces  in this region are infinite when the singularity is gravitationally strong. When these forces have finite  real value or equal to zero then the central singularity is gravitationally weak.

\setcounter{equation}{0}
\section{5.Conclusion}

In this paper we have considered Vaidya spacetime when the mass function is $M(v)=\lambda v^n$.  We found out that when $n\geq 1$ the result of the gravitational collapse might be the naked singularity. It means that there exist a family of non-spacelike future-directed geodesics which terminate in the central singularity in the past and the time of the apparent horizon is bigger than the time of the singularity formation. However if there is such a family of geodesics then particles can move along them and we  investigated the question which inertial forces act on this particle when it escapes the naked singularity. We found out that these inertial forces are infinite when $n \in [1\,, 2)$ and finite when $n\geq 2$. However the naked singularity is the gravitationally strong only in the case when $n=1$. When $n>1$ it is gravitationally weak. So we can conclude that the inertial forces which act on the particle when it escapes the singularity are finite only when the singularity is gravitationally weak and infinity when it is gravitationally strong. However, the weak naked singularity doesn't have much interest because the manifold can be extended through it. Thus if we consider the black hole or naked singularity formation in the case of Vaidya spacetime one should consider the linear mass function $M(v)=\lambda v$, where $\lambda$ is real positive constant. One should also note that only the linear mass function is the solution of the Einstein equation which violate  CCC because it states that only the strong singularity must be always covered with horizon. 

\textbf{Acknowledgments} The work was performed within the SAO RAS state assignment in the part "Conducting Fundamental Science Research".

\end{document}